\documentclass[12pt]{article}

\usepackage{amssymb,amsmath}
\usepackage{cite}

\newcommand{\dd}{\partial}
\newcommand{\de}{\delta}
\newcommand{\m}{\mu}
\newcommand{\n}{\nu}
\newcommand{\ls}{\left(}
\newcommand{\rs}{\right)}
\newcommand{\al}{\alpha}
\newcommand{\la}{\lambda}
\newcommand{\tr}[1]{\overset{{\scriptscriptstyle 3}}{#1}{}}
\newcommand{\pos}{\tr{\Pi}_{\!\!\bot}{}}
\newcommand{\pua}[2]{\left\{#1,#2\right\}}
\newcommand{\dz}{\zeta}
\newcommand{\na}{\nabla\!}

\newcommand{\disn}[2]{$$\displaylines{\refstepcounter{equation}%
            \label{#1}\hskip 1em minus 1em #2\hfilneg}$$}
\newcommand{\nom}{\hfil\hskip 1em minus 1em (\theequation)}
\newcommand{\no}{\hfil \hskip 1em minus 1em\phantom{(\theequation)}%
            \hfilneg\cr\hfilneg\hskip 1em minus 1em\hfil}
\newcommand{\ns}{\hfill\cr\hfill}

\begin{document}

\title{External time canonical formalism for gravity\\
in terms of embedding theory}
\author{
S.A.~Paston\thanks{E-mail: paston@pobox.spbu.ru},
E.N.~Semenova\thanks{E-mail: derenovacio@gmail.com}\\
{\it Saint Petersburg State University, Saint Petersburg, Russia}
}
\date{\vskip 15mm}
\maketitle

\begin{abstract}
The formulation of gravity theory is considered where space-time is a 4-dimensional surface
in flat ten-dimensional space. The possibility of using the "external" time (the time of ambient space) in
this approach is investigated. The transition to the "external" time is realized with the help of partial gauge
fixing, the coordinate condition which equates the timelike coordinate of the surface and time of the ambient
space. It is shown that by using such a gauge condition in the action, the loss of any equations of motion
does not take place, although it can happen in the general case. A version of the canonical formalism
of the theory is studied in which certain additional constraints are imposed, providing the equivalence of
the approach under consideration and general relativity. The corresponding first-class constraint algebra
is obtained. It is proved that using the gauge directly in the action leads to the same result as gauge
fixing in the constraint algebra, despite the artificial introduction of some of the constraints into the theory.
Application of the "external" time may be useful for attempts to quantize the theory as well as for studies
in embedding theory where 4-dimensional surfaces fill the whole ambient space and coordinates on the
surfaces are not introduced.
\end{abstract}

\newpage

\section{Introduction}
Embedding theory is an alternative approach in describing gravity. In this approach, our 4-dimensional space-time is not considered as an abstract Riemannian space but as a 4-dimensional surface in flat ambient space of larger dimension (multidimensional Minkowski space). Then the dynamic variables describing the gravitational field are not the components of the metric $g_{\m\n}(x^\al)$ but the embedding function $y^a(x^\al)$ that determines the surface (here and henceforth $\m,\n,\ldots=0,1,2,3$, $a,b,\ldots=0,1,\ldots,N-1$,where $N$ is the number of ambient space dimensions).
According to Friedmann's theorem which generalizes the Janet-Cartan theorem to pseudo-Euclidean spaces, for a space of signature $(1,3)$ with an arbitrary analytic metric to be locally represented as such a surface, one should take $N\ge10$, and the signature of ambient space-time must have the form $(r,N-r)$ with $1\le r\le N-3$ (see, e.g., \cite{goenner}). Thus, without restricting the generality from a local viewpoint, it is sufficient to use an ambient space with $N=10$ and $r=1$, i.e., with one timelike direction (we will use the signature $+-\ldots-$).

Such a choice is preferable from a physical viewpoint because the existence of a few timelike directions complicates interpretation of the results of the theory, especially if we further pass on to a certain field theory in the ambient space \cite{statja25}. One should, however, note that when constructing explicit embeddings of physically interesting solutions to the Einstein equations, one also uses some versions of the ambient space with $r>1$,  i.e., with a few timelike directions. This can happen, in particular, if a certain symmetry of the surface is assumed (see examples in \cite{statja27}) or if a global nature of the embedding is required, i.e., its smoothness at all points including the horizons (see examples in \cite{statja30}).

The metric is expressed in terms of the embedding function according to the induced metric formula
 \disn{1}{
g_{\m\n}(x)=\dd_\m y^a(x)\dd_\n y^b(x)\eta_{ab},
\nom}
where $\eta_{ab}$ is the flat metric of the ambient space, see~\cite{goenner}.

For the first time, a description of gravity, similar to the approach of string theory, in the form of a theory of a 4-dimensional surface in 10-dimensional space with the signature (1,9) was suggested in \cite{regge} and discussed in \cite{deser}.Later on, the idea of embedding in flat space was repeatedly used for the description of gravity \cite{pavsic85let,tapia,maia89,bandos,estabrook1999,davkar,bustamante,statja18}. It is of interest to note that in \cite{pavsic85let,estabrook1999}  this idea has emerged independently. Studies in the embedding theory are also being continued in the recent years \cite{faddeev,statja26,willison}. It is worth noting that the approach of embedding theory is essentially different from brane theory which has recently become popular \cite{lrr-2010-5} in that in embedding theory the ambient space is flat and does not contain gravity.

In the framework of the idea of space-time embedding in flat ambient space, there are also attempts to connect the quantum effects in space-time and in the ambient space \cite{deserlev99,banerjee10,statja34} (see also the list of references in \cite{statja34}), and for this purpose explicit embeddings of physically relevant solutions to the Einstein equations are built, see, e.g., \cite{collinson68,davidson,statja27,statja30} as well as references in \cite{statja30}. A detailed, although somewhat obsolete list of references connected with embedding theory and related issues can be found in the review paper \cite{tapiaob}. A formalism convenient for making calculations in embedding theory is presented in detail in the book \cite{mon1}.

A description of gravity in the framework of embedding theory may turn out to be useful in attempts to build a quantum theory of gravity because in this approach a flat space emerges in a natural way. Many problems appearing in attempts of quantization of gravity in terms of the metric (presented, for example, in the review \cite{carlip}), emerge because we try to apply the quantization procedure, which has been quite successful in field theory in flat space, to the case where the dynamic variables are the geometric characteristics of space and it is these quantities that should be quantized.

In particular, one of the important problems is the problem of choosing time. Since in the framework of a general covariance symmetry the theory is invariant under time reparametrization, the Hamiltonian of the theory reduces to a linear combination of the constraints (see the Arnovitt-Deser-Misner (ADM) formalism \cite{adm}), and, as a result, in the Schr{\"o}dinger picture the state vector does not change with time. It is usually supposed that this indicates that the role of "physical" time should not be played by the coordinate $x^0$but some other quantity, however, a satisfactory method of building such a quantity in the framework of GR is unknown in the general case. Since embedding theory is formulated in Minkowski ambient space, its timelike coordinate $y^0$can play such a role in this formulation of gravity.

One more problem is a formulation of the causality principle in gravitation theory. In quantum field theory the causality principle usually means that the operators corresponding to regions separated by a spacelike interval should commute. In the usual formulation of gravitation in terms of the metric $g_{\m\n}$,it is hard to formulate such a principle because the interval between points is determined by the metric which itself is an operator. Therefore, for a given pair of space-time points, it is not possible in an absolute sense, i.e., independently of the state vector, to say by which kind of interval they are separated.  However, if one describes gravity in the framework of embedding theory, one can build a variant of the theory in the form of a field theory in ambient space, and then causality can be defined in the usual way for quantum field theory. Such a variant of embedding theory has been suggested in \cite{statja25}. In this "splitting theory", a field specified in the ambient space describes a set of noninteracting 4-dimensional surfaces, and each of these obeys the same equations as in embedding theory and can be considered as our space-time.

To perform canonical quantization of the theory, it is necessary to formulate it in a canonical (Hamiltonian) form. For embedding theory (and therefore for the splitting theory) it turns out to be a complicated problem. If one takes, as is usually done in embedding theory, the action in the Einstein-Hilbert form, substituting there the metric in the form (\ref{1}), then it will contain time derivatives of higher than the first order. The Hamiltonian formulation in this case requires using some special methods, and this approach was developed in \cite{tapia,rojas06}. However, one can notice that, rejecting the surface terms in the action, one can bring it to a form which does not contain higher than the first-order time derivatives. In this case, the canonical formalism is developed in the usual way, but either some of the constraints cannot be written in an explicit form \cite{frtap}, or they can only be written by introducing additional variables, which leads to the emergence of second-class constraints \cite{davkar}. Some results in quantization of embedding theory have been obtained in \cite{pavsic85,davkar}.

The opportunity to simplify the canonical formalism of embedding theory, simultaneously solving the problem of extra solutions
(the equations of embedding theory can lead to some other,
"extra" solutions in addition to those of the Einstein equations,
they are discussed in \cite{davids97,statja18,statja26,statja33})
was noticed as early as in \cite{regge}.
To this end it is necessary, in addition to the constraints emerging in the canonical formalism, to impose {\it ad hoc} four more constraints, the socalled "Einsteinian" constraints (4 of the 10 Einstein equations). As a result, there emerges the Regge-Teitelboim formulation of gravitation which turns out to be equivalent to Einstein's. The canonical formalism of this theory with "internal" time $x^0$ was built in \cite{statja18,statja24}. It was found that the Hamiltonian of the theory reduces to a combination of eight constraints (four of which are the additionally imposed "Einsteinian" constraints), and for these constraints, the corresponding algebra of first-class constraints was built. A relationship was also revealed between this algebra and the algebra describing general relativity (GR) in the ADM approach \cite{adm}.

The purpose of the present paper is a study of the "external" time ($y^0$) canonical formalism for formulation of the Regge-Teitelboim gravitation theory. To that end, it is necessary to pass over from the description of the surface by the embedding function $y^a(x^\al)$ to its description by the function $y^A(y^0,x^i)$ (here and henceforth $i,k,\ldots=1,2,3$ and $A,B,\ldots=1,\ldots,9$), specifying the dynamics by the time $y^0$ of the threedimensional section of our space-time. This transition
reduces to imposing the condition
 \disn{2}{
y^0(x^\al)=x^0,
\nom}
which partly fixes the generally covariant arbitrariness of the coordinate choice on the surface, which is partial gauge fixing. The transition from the "internal" time $x^0$ chosen arbitrarily to the "external" time $y^0$ of Minkowski ambient space seems natural from the viewpoint of time choice at quantization. Besides, such a transition turns out to be an intermediate step on the way to the construction of the canonical formalism for the above-mentioned splitting theory suggested in \cite{statja25}. This is the case because the splitting theory describes the set of surfaces, each of these corresponding to the single surface of embedding theory but describes them in a coordinate-free manner (the coordinates $x^\m$ on the surfaces are not introduced), so that in the canonical formalism one can use only the time of ambient space.

In Section 2 of the present paper we consider examples of simple theories in which gauge fixing
in the action can both change and leave unchanged the equations of motion; we show how the canonical
descriptions of the full theories and theories with a fixed gauge are connected. In Section 3 we study the
influence of using the condition (\ref{2}) in the action on the equations of motion in embedding theory. Section
4 is devoted to the construction of a canonical description of the Regge-Teitelboim formulation with
"external" time, while in Section 5 we calculate the constraint algebra of such a theory. In Section 6
the results obtained are compared with the canonical formalism with "internal" time obtained in \cite{statja18,statja24}.
It is not easy to establish a relationship between these two formalisms a priori without performing complete
calculations because they are not constructed in a totally standard way from the action, and in the process
of construction, the Einsteinian constraints are additionally imposed.

\section{Examples of gauge fixing in the action}
\subsection{A gauge that changes the theory}
Consider an example of a theory in which gauge fixing in the action changes the equations of motion.
Consider the action of free electrodynamics
\disn{p1.1}{
S=-\frac{1}{4}\int\! d^4x\; F_{\m\n}F^{\m\n}=\int\! d^4x\;\ls \frac{1}{2}F_{0i}F_{0i}-\frac{1}{4}F_{ik}F_{ik}\rs,\qquad
F_{\m\n}=\dd_\m A_\n-\dd_\n A_\m.
\nom}
The equations of motion have the form
\disn{p1.2}{
\dd_\m F^{\m\n}=0.
\nom}
Note that these four equations are connected by the differential identity $\dd_\n\dd_\m F^{\m\n}=0$.
Let us write down the canonical formalism which is well known for this theory. We have the generalized momenta $\pi_i=F_{0i}$, $\pi_0=0$, which create the primary constraint $\phi_1=\pi_0\approx0$ (here and in what follows the sign $\approx$ denotes
a "weak" equality which cannot be used before calculating all Poisson brackets). Having written the
expression for the Hamiltonian and finding the secondary constraint $\phi_2=\dd_i\pi_i\approx0$, we finally find the
generalized Hamiltonian of the theory in the form
\disn{p1.3}{
H=\int\! d^3x\ls\frac{1}{2}\pi_i\pi_i+\frac{1}{4}F_{ik}F_{ik}+\la_1\phi_1+\la_2\phi_2\rs,
\nom}
where $\la_{1,2}$ are the Lagrange multipliers. The constraints $\phi_1$ and $\phi_2$ form the system of first-class constraints
since their Poisson brackets are zero.

Now let us use the gauge $A_0=0$ in the action, after which it takes the form
\disn{p1.4}{
S=\int\! d^4x\;\ls \frac{1}{2}(\dd_0 A_i)(\dd_0 A_i)-\frac{1}{4}F_{ik}F_{ik}\rs.
\nom}
Varying it in the remaining variables $A_i$, we obtain the new equations of motion
\disn{p1.5}{
\dd_\m F^{\m i}=0,
\nom}
and their number is smaller that in (\ref{p1.2}) by one.
Thus it is evident that in this example gauge fixing in the action leads to a nonequivalent theory in which one
of the equations has been lost. One can note that because of the differential identity that connects equations
(\ref{p1.2}) the lost equation reduces to a condition for the initial data, which, having been imposed, is
conserved due to the remaining equations of motion (see an analysis of this example in \cite{statja24}), nevertheless,
the theory (\ref{p1.4}) turns out to be nonequivalent to the theory (\ref{p1.1}).

Let us write down the canonical formalism for the theory (\ref{p1.4}). For the generalized momentum we
have $\pi_i=\dd_0A_i$, constraints are absent, and for the Hamiltonian of the new theory we find
\disn{p1.6}{
H_{\text{new}}=\int\! d^3x\ls\frac{1}{2}\pi_i\pi_i+\frac{1}{4}F_{ik}F_{ik}\rs.
\nom}
Let us now look what happens if in the full theory (\ref{p1.1}) now in the Hamiltonian formalism, we fix
the gauge by introducing it as the additional condition $\chi=A_0\approx0$. Since the Poisson bracket of the constraints $\{\phi_1,\chi\}\ne0$, these are second-class constraints, and they can be solved by expressing with
their aid the pair of conjugated variables $A_0,\pi_0$ in terms (in the general case) of other variables in the
first-order Lagrangian \cite{popov}:
\disn{p1.7}{
L^{(1)}=\left.\ls\int\! d^3x\ls\pi_0\dd_0 A_0+\pi_i\dd_0 A_i\rs-H\rs\right|_{A_0=0,\pi_0=0}=\ns=
\int\! d^3x\ls\pi_i\dd_0 A_i-\ls\frac{1}{2}\pi_i\pi_i+\frac{1}{4}F_{ik}F_{ik}+\la_2\phi_2\rs\rs,
\nom}
whence we find the expression for the Hamiltonian of the full theory (\ref{p1.1}) with the eliminated pair of variables:
\disn{p1.8}{
\tilde H=\int\! d^3x\ls\frac{1}{2}\pi_i\pi_i+\frac{1}{4}F_{ik}F_{ik}+\la_2\phi_2\rs.
\nom}
As we see, in the present case the Hamiltonian describing the full theory does not coincide with the
Hamiltonian of the new theory (\ref{p1.6}), and the difference consists in the term containing the constraint $\phi_2$ with the Lagrange multiplier. It is this constraint that forms the lost equation of motion that distinguishes (\ref{p1.2}) from (\ref{p1.5}).

The above example shows that gauge fixing in the action leads in the general case to a changed
theory, unlike gauge fixing in the equations of motion or in the Hamiltonian formalism by introduction
of additional conditions. However, in some cases gauge fixing in the action can also leave the
theory unchanged, and the corresponding example is considered in Section 2.2.

\subsection{A gauge that does not change the theory}
Let us now consider an example of a theory in which gauge fixing in the action does not change
the equations of motion. Consider the action of a relativistic particle
\disn{p2.1}{
S=-\int\! d\tau \sqrt{\dot x^\m(\tau) \dot x_\m(\tau)},\qquad
\dot x^\m(\tau)\equiv\frac{d}{d\tau}x^\m(\tau).
\nom}
We note that this theory has much in common with the description of gravity in embedding theory. The
equations of motion due to the action (\ref{p2.1}) have the form
\disn{p2.2}{
\dot u^\m=0,\qquad
u^\m\equiv \frac{\dot x^\m}{\sqrt{\dot x^\n\dot x_\n}}.
\nom}
These four equations are connected by the algebraic identity $u_\m\dot u^\m=0$. Let us write down the canonical formalism, well known for this theory. We have the generalized momentum $p_\m=-u_\m$, which creates the
primary constraint $\phi=p_\m p^\m-1\approx0$. It is easy to
verify that the Hamiltonian for this theory turns out to be zero as it is for any theory symmetric with
respect to time reparametrization, while the generalized Hamiltonian reduces to a term containing a
constraint with the Lagrange multiplier:
\disn{p2.3}{
H=\la\phi.
\nom}

Let us now use a gauge in the action: the coordinate condition  $x^0(\tau)=\tau$, which identifies the particle's time ("internal" time, compare with (\ref{2})) with the time of Minkowski space ("external" time). As a result, the action takes the form
\disn{p2.4}{
S=-\int\! d\tau \sqrt{1-\dot x^i(\tau) \dot x^i(\tau)}.
\nom}
Varying it in the remaining variables $x^i$, we obtain the new equations of motion
\disn{p2.5}{
\frac{d}{d\tau}\ls\frac{\dot x^i}{\sqrt{1-\dot x^k\dot x^k}}\rs=0,
\nom}
which coincide with three of the four equations (\ref{p2.2}),while the fourth equation has been lost. However, in
this case, unlike the one considered in Sec. 2.1, because of the algebraic identity that connects Eqs. (\ref{p2.2}) and the condition $u^0\ne0$ this lost equation turns out to follow from the other three. Indeed, from (\ref{p2.5}) it
follows that the quantity $\dot x^k\dot x^k$ does not depend on $\tau$, hence immediately follows the validity of Eq. (\ref{p2.2}) at $\m=0$.

Thus, in this example, gauge fixing in the action leaves the theory unchanged, leading to the same
equations of motion. This fact can also be explained from a geometric viewpoint. Indeed, the equations of
motion follow from the requirement that variation of the action vanishes at small variations $\de x^\m(\tau)$ of the
independent variable. However, since the form of the action (\ref{p2.1}) requires that the vector $u^\m$,
tangent to the particle world line should be timelike at each point, an arbitrary small deformation of the particle world
line may be provided by a variation $\de x^\m(\tau)$, in which $\de x^0(\tau)=0$. It is therefore unimportant whether the
variable $x^0(\tau)$ is varied or not varied due to gauge fixing in the action. Precisely the same situation
takes place in the description of gravitation in the framework of embedding theory, whose equations of
motion are analyzed in Section 3.

Consider the well-known canonical formalism for the theory (\ref{p2.4}). Writing down the expression for the
momentum $p_i$ and making sure that there emerge no constraints, it is easy to obtain the expression for the Hamiltonian
\disn{p2.6}{
H_{\text{new}}=\sqrt{1+p_ip_i}.
\nom}
Let us now look what happens if, for the full theory (\ref{p2.1}), now in the Hamiltonian formalism we fix the
gauge by introducing it as the additional condition $\chi=x^0-\tau\approx0$. Since the Poisson bracket of the
constraints $\{\phi,\chi\}\ne0$, these are second-class constraints, and they can be solved by expressing, with
their aid, the pair of conjugated variables $x^0,p_0$ in terms of other variables in the first-order Lagrangian
(herewith for $p_0$ one should take the negative value because
$p_0=-u^0<0$ at $x^0=\tau$):
\disn{p2.7}{
L^{(1)}=\left.\ls p_0\dot x^0+p_i\dot x^i-H\rs\right|_{x^0=\tau,p_0=-\sqrt{1+p_i p_i}}=
p_i\dot x^i-\sqrt{1+p_i p_i}=p_i\dot x^i-\tilde H,
\nom}
from which we find that the expression for the Hamiltonian $\tilde H$ of the full theory (\ref{p2.1}) with the excluded pair of variables coincides with the Hamiltonian of the new theory (\ref{p2.6}).

Thus in the above example the canonical formalism emerging after gauge fixing in the action can be
obtained as a result of gauge fixing in the canonical formalism of the full theory and by introduction of
an additional condition. For the Regge-Teitelboim formulation of gravity, one cannot be sure about a
similar result because it does not directly follow from the action but emerges after additionally imposing the
Einsteinian constraints in the process of building the canonical formalism.

\section{The equations of motion using "external" time}
Let us find out whether the equations of motion of embedding theory will change if we use the "external"
time as time in our dynamics, i.e., impose the condition (\ref{2})in the action. As the action of the embedding theory, we take, as is usually done, the Einstein-Hilbert action
\disn{p3.1}{
S=\int d^4x\sqrt{-g}R,
\nom}
in which we substitute the induced metric (\ref{1}). Varying this action with respect to the embedding function $y^a(x)$, we can obtain the equations
of motion of the theory, the Regge-Teitelboim equations \cite{regge}, having the form
 \disn{p3.2}{
G^{\m\n}\,b^a_{\m\n}=0,
\nom}
where $G^{\m\n}$ is the Einstein tensor, $b^a_{\m\n}=\na_\m \na_\n\, y^a$ is the second fundamental form of the surface, and $\na_\m$ is the covariant derivative (see the details about the formalism of the embedding theory and the notations
used in \cite{statja18,statja24}). Note that Eqs. (\ref{p3.2}) can contain, besides the solutions to the Einstein equations, some
"extra" solutions, they are discussed in \cite{davids97,statja18,statja26,statja33}).

In Section 2 we have shown that gauge fixing in the action may both change the equations of motion
and leave them unchanged. Let us find out which of these variants is realized when using the condition (\ref{2})
in the action of the theory (\ref{p3.1}), that is, whether or not Eqs. (\ref{p3.2}).
will change in this case. Equating to zero the variation of the action under an arbitrary variation
of $y^A(x)$ and under the condition (\ref{2}):
\disn{l21}{
\de S=-\int d^4x\;\sqrt{-g}\;G^{\m\n}\de g_{\m\n}=
-\int d^4x\;\sqrt{-g}\;G^{\m\n}\ls e_{\m, A}\dd_{\n}\de y^A+e_{\n, A}\dd_{\m}\de y^A\rs=0
\nom}
(here $e_{\m,A}=\dd_\m y_A$),
one can obtain the equations of motion in the theory with "external" time in the form
\disn{l20}{
G^{\m \n}b^A_{\m \n}=0.
\nom}
These are nine of the ten Regge-Teitelboim equations (\ref{p3.2}).

At first sight, the theory with "external" time turns out to be nonequivalent to the original one because
one of the equations is lacking. However, it is not the case since the lacking equation
\disn{l23}{
G^{\m\n}\; b^{(0)}_{\m\n}=0
\nom}
(the index "(0)" is here written in parentheses to designate that it is the zeroth component of a 10-
dimensional rather than 4-dimensional quantity) follows from the other equations (\ref{l20}).
This can be proved by contradiction. Let Eqs. (\ref{l20}) be satisfied but not Eq. (\ref{l23}). Then the vector $u^a\equiv G^{\m\n}b^a_{\m\n}$ has only the zeroth component, hence it is timelike. However,
it is known (see, e.g., \cite{statja18}), that for any vector $m^a$ tangent to a surface, holds the equation $m_a b^a_{\m\n}=0$, consequently, $m_a u^a=0$,  i.e., the vector $u^a$ is orthogonal
to the surface. Since we assume that the ambient space contains only one timelike direction, and at
each point of the surface there is a timelike vector tangent to it, we arrive at a contradiction.

Thus we obtain that the equations of motion of the theory with "external" time are equivalent to the
Regge-Teitelboim equations, whose derivation used an arbitrary choice of time. The same result could also
be obtained from geometric considerations. Indeed, the surface has a timelike tangent vector at its every
point, therefore its arbitrary small deformations can be specified by moving it along spacelike directions
only. Consequently, increments of the action at arbitrary variations of $y^a$
reduce to its increments at variations of only the components of $y^A$, and the equations of motion coincide.

\section{The canonical formalism in "external" time}
In the previous section we have seen that using the "external" time does not change the equations
of motion of the theory. Let us now investigate the canonical formalism in "external" time for the Regge-
Teitelboim formulation of gravitation, i.e., when the Einsteinian constraints are additionally imposed. To
do that, we impose the condition (\ref{2}) in the action written in the form obtained in \cite{statja24} in which the
dependence on $\dot y^a\equiv \dd_0 y^a$ is explicit. Let us note that, under the condition (\ref{2}), the threedimensional
spacelike submanifold $x^0=const$ of our space-time turns out to be embedded in the nine-dimensional
flat space $y^{0}=const$. Therefore, for the second fundamental form $\tr{b}^a_{ik}$ of this submanifold (all
quantities corresponding to it will be marked by "3") the relation $\tr{b}^{(0)}_{ik}=0$ is valid, while for the projector $\pos_{ab}$ to the space transversal to this submanifold at a given point, hold the relations $\pos^{(0)(0)}=1$ and $\pos^{(0)A}=0$. Using these facts, it is not hard to obtain the expression for the action in the form
\disn{l9}{
S=\int dx^0 L(y^a,\dot y^a),\quad
L=\int d^3x\;\frac{1}{2}\ls
\frac{\dot y^A\;B_{AB}\;\dot y^B}{\sqrt{1+\dot y^A\;\pos_{AB}\;\dot y^B}}+
\sqrt{1+\dot y^A\;\pos_{AB}\;\dot y^B}\;B^D_D\rs,
\nom}
where
\disn{l5a}{
B^{AB}=2\sqrt{-\tr{g}}\;\tr{b}^A_{ik}\tr{b}^B_{lm}L^{ik,lm},\qquad
L^{ik,lm}=\tr{g}^{ik}\tr{g}^{lm}-\frac{1}{2}\ls\tr{g}^{il}\tr{g}^{km}+\tr{g}^{im}\tr{g}^{kl}\rs.
\nom}
The quantity $L^{ik,lm}$ used here coincides, up to a factor, with theWheeler-DeWitt superspace metric.

Let us find the generalized momentum $\pi_A$ for the variable $y^A$ from the action (\ref{l9}):
\disn{l12}{
\pi_A=\frac{\de L}{\de \dot y^A}=B_{AF}n^F+\frac{1}{2}n_A\ls B^D_D-n^FB_{FH}n^H\rs,
\nom}
where
\disn{l11}{
n^A=\frac{\pos_B^A\;\dot y^B}{\sqrt{1+\dot y^A\;\pos_{AB}\;\dot y^B}}.
\nom}
Taking into account (\ref{l5a}) and the properties of the quantity $\tr{b}^A_{ik}$, from the relation
(\ref{l12}) we obtain three primary constraints
\disn{p4.10}{
\tilde \Phi_i=\pi_A\tr{e}^A_i=0,
\nom}
where $\tr{e}^A_i=\dd_i\tr{y}^A$.
Let us suppose that, in addition to the constraints emerging in the usual way, there must
be additionally imposed four Einsteinian constraints
\disn{l60}{
n_{\m}G^{\m\n}\approx0,
\nom}
where $n_{\m}$ is the unit normal to the submanifold $x^0=const$, for which part of the components of the corresponding
vector in the ambient space are specified by Eq. (\ref{l11}). Applying the condition (\ref{2}) to the expressions
for these constraints presented in \cite{statja24}, we can write them down in the form
\disn{l61}{
{\cal\tilde H}^0=\frac{1}{2}\ls n_A B^{AD} n_D-B^A_A\rs\approx0,\qquad
{\cal\tilde H}^i=-2\sqrt{-\tr{g}}\;\,\tr{\na}_k\ls L^{ik,lm}
\,\tr{b}^A_{lm}\, n_A\rs\approx0.
\nom}
If we use the constraint ${\cal\tilde H}^0$ in the relation (\ref{l12}), it takes the simple form
\disn{p4.9}{
\pi_A=B_{AF}n^F.
\nom}

The quantity $n^A$ is unambiguously related to the transversal component of the velocity $\pos^A_B\dot y^B$:
\disn{l103.1}{
\pos^A_B\dot y^B=\frac{n^A}{\sqrt{1-n^An_A}},
\nom}
since the square of $n^A$ can be expressed in terms of $\dot y^A$ by the relation
\disn{l102}{
n^An_A=\frac{\dot y^A\;\pos_{AB}\;\dot y^B}{1+\dot y^A\;\pos_{AB}\;\dot y^B}.
\nom}
The opportunity to express the transversal component of the velocity unambiguously in terms of $n^F$,
and hence in terms of the momentum $\pi_A$, makes a difference of principle between the theory with a partly
fixed gauge (\ref{2}) from the approach without such fixing, because in the usual approach the vector $n^a$ is
normalized to unity, and this leads to the emergence of one more primary constraint, see~\cite{statja24}, which does
not emerge in the case under consideration here.

Using Eqs. (\ref{l11}) and (\ref{p4.9}), it is easy to find that the Hamiltonian of the theory
\disn{l13}{
H=\int d^3x\,\pi_A\dot y^A-L
\nom}
turns out to be proportional to the constraint ${\cal\tilde H}^0$, hence it is equal to zero in the weak sense. Therefore
the generalized Hamiltonian reduces to a linear combination of the three constraints (\ref{p4.10}) and the four
constraints (\ref{l61}). It turns out to be convenient to consider, instead of the constraints ${\cal\tilde H}^i$, the linear combination $\tilde\Psi^k={\cal\tilde H}^{k}+\tilde\Phi_{i} \tr g^{ik}$. It will be shown below
(see Section 5) that, as in the approach without using the condition (\ref{2}), such a linear combination generates
the transformations which are isometric bendings of the submanifold $x^0=const$.
Now the set of seven constraints will look as follows:
$\tilde\Phi_i, \tilde\Psi^i$ and ${\cal\tilde H}^0$.

In the canonical formalism, the constraints should be expressed in terms of generalized coordinates and
momenta. To bring them to the necessary form, we introduce a quantity inverse with respect to $\tr b^A_{lm}$ in
the following sense (note that the quantity $\pos^B_A$ is a projector onto a 6-dimensional space):
\disn{l26}{
\tr b^{-1\;ik}_{\;\;\;\;\;A}\;\tr b^A_{lm}=\frac{1}{2}\ls\de^i_l\de^k_m+\de^i_m\de^k_l\rs,\qquad
\tr b^{-1\;ik}_{\;\;\;\;\;A}\;\tr b^B_{ik}=\pos^B_A\;,\qquad
\tr b^{-1\;ik}_{\;\;\;\;\;A}=\tr b^{-1\;ki}_{\;\;\;\;\;A},
\nom}
as well as a quantity inverse to $L^{ik,lm}$:
\disn{l33}{
 \hat L_{pr,ik}\;L^{ik,lm}=\frac{1}{2}\ls\de^l_p\de^m_r+\de^l_r\de^m_p\rs, \qquad
\hat L_{lm,ik}=\frac{1}{2}\ls\tr g_{lm}\tr g_{ik}-\tr g_{li}\tr g_{mk}-\tr g_{lk}\tr g_{mi}\rs.
\nom}
As a result, we obtain the set of seven constraints
\disn{l25}{
\tilde\Phi_i=\pi_A\tr{e}^A_i,\qquad
{\tilde\Psi}^i=-\sqrt{-\tr g}\;\tr \na_k\ls\frac{1}{\sqrt{-\tr g}}\tr b^{-1\;ik}_{\;\;\;\;\;A}\;\pi^A\rs+\pi^A\tr{e}^i_A,
\no
{\cal\tilde H}^0=\frac{1}{4\sqrt{-\tr{g}}}\;\pi^A\tr b^{-1\;ik}_{\;\;\;\;\;A} \hat L_{ik,lm}\tr b^{-1\;lm}_{\;\;\;\;\;B}\pi^B-\sqrt{-\tr{g}}\;\tr b^A_{ik}\tr b^D_{lm}\eta_{DA}L^{ik,lm},
\nom}
and the generalized Hamiltonian of the theory reduces to their linear combination:
\disn{l25.1}{
\tilde H^{\text{gen}}=\int d^3x\ls
\la^i\tilde\Phi_i+N_i{\tilde\Psi}^i+N_0{\cal\tilde H}^0
\rs
\nom}
using the "external" time. Note that without using the condition (\ref{2}) there are eight constraints (see~\cite{statja24}).

\section{The algebra of constraints when using\\ the "external" time}
To verify the closure of the algebra of constraints and to find its precise form, it is necessary to calculate
the Poisson brackets of the constraints (\ref{l25}) with each other. To do that, it is convenient to pass on to convolutions
of the constraints with arbitrary functions $\xi^i(x) and \xi(x)$, introducing the quantities
\disn{l27}{
\tilde\Phi_\xi\equiv\int d^3x\; \tilde\Phi_i(x)\,\xi^i(x)=\int d^3x\;\pi_A\tr{e}^A_i\xi^i,\qquad
{\cal\tilde H}^0_\xi\equiv\int d^3x\; {\cal\tilde H}^0(x)\,\xi(x),\no
{\tilde\Psi}_\xi\equiv\int d^3x\; \tilde\Psi^i(x)\xi_i(x)=\int d^3x\;\pi^A\ls\tr b^{-1\;ik}_{\;\;\;\;\;A}\;\tr \na_i\xi_k+\tr{e}_A^i\xi_i\rs.
\nom}
Their usage simplifies the calculations and makes it possible to write down the results in a more compact
form. The calculations necessary for obtaining the Poisson brackets of the seven constraints (\ref{l27}) with
each other to a large extent repeat those performed in \cite{statja24}, therefore we here present them in rather a brief
form.

First of all, let us make clear the geometric meaning of the three constraints $\tilde\Phi_i$, and to do so, we
calculate their action on the dynamic variables:
\disn{l31}{
\pua{\tilde\Phi_\xi}{y^A(x)}=\xi^i(x)\dd_i y^A(x),\qquad
\pua{\tilde\Phi_\xi}{\frac{\pi_A(x)}{\sqrt{-\tr{g}(x)}}}=\xi^i(x)\dd_i \frac{\pi_A(x)}{\sqrt{-\tr{g}(x)}},
\nom}
where $\{\dots\}$ are Poisson brackets. This means that $\tilde\Phi_\xi$ generates the transformation $x^i\to x^i+\xi^i(x)$ of
the three-dimensional coordinates on the surface of constant time (it should be noted that the generalized
momentum $\pi_A(x)$ is a 3-dimensional scalar density). The same is also true in the approach without using
the condition (\ref{2}), see~\cite{statja24}. Since all the constraints (\ref{l25}) are tensor densities, one can immediately
write down their action on the corresponding convolutions (\ref{l27}) of the constraints $\tilde\Phi_i$:
\disn{g5}{
\pua{\tilde\Phi_\xi}{\tilde\Phi_\dz}=-\int d^3x\;\tilde\Phi_k\ls\xi^i\tr \na_i\dz^k-\dz^i\tr \na_i\xi^k\rs,
\nom}
\vskip -1em
\disn{q4.1}{
\pua{\tilde\Phi_\xi}{\tilde\Psi_\dz}=-\int d^3x\;\tilde\Psi^k\ls\xi^i\tr \na_i\dz_k+\dz_i\tr \na_k\xi^i\rs,
\nom}
\vskip -1em
\disn{g5.2}{
\pua{\tilde\Phi_\xi}{{\cal\tilde H}^0_\dz}=-\int d^3x{\cal\tilde H}^0\xi^i\dd_i\dz.
\nom}

Further, let us make clear the geometric meaning
of the three constraints $\tilde\Psi^i$. It turns out that
 \disn{g6}{
\pua{\tilde\Psi_\xi}{\tr{g}_{ik}(x)}=0.
\nom}
This means that $\tilde\Psi_\xi$ generates isometric bendings of the 3-dimensional constant-time submanifold.
Since, due to the condition (\ref{2}) this submanifold is embedded in flat 9-dimensional space, the number
of the existing bending generators is three, which corresponds to the difference between the number of
ambient space dimensions and the number of components in the 3-dimensional metric, i.e., six. Note that
without using the condition (\ref{2}), the ambient space is 10-dimensional, and the number of generators of
bendings becomes four, see~\cite{statja24}. Using (\ref{g6}), after rather cumbersome calculations it turns out to be
possible to find the result of action of the constraint $\tilde\Psi_\xi$ on itself and on ${\cal\tilde H}^0_\dz$:
\disn{f2}{
\pua{\tilde\Psi_\xi}{\tilde\Psi_\dz}=\int d^3x\, \tilde\Psi^i\ls  r_{\xi A} \tr \na_i r^A_\dz-
  r_{\dz A} \tr \na_i r^A_\xi\rs,
\nom}
\vskip -1em
\disn{l107}{
\pua{\tilde\Psi_\xi}{{\cal\tilde H}^0_\dz}=\int d^3x\, \tilde\Psi^i\ls  r^C_\xi \tr{\na}_i\ls\pi^B\tilde{B}_{BC}\dz\rs-
\ls\tr{\na}_i r^C_\xi\rs\pi^B\tilde{B}_{BC}\dz\rs,
\nom}
where
\disn{l103}{
r^B_\xi\equiv\frac{\de\tilde\Psi_\xi}{\de\pi_B}=\tr b^{-1\;ik\;B}\tr \na_i\xi_k+\tr e^{iB}\xi_i,\qquad
\tilde{B}_{EC}=\frac{1}{2\sqrt{-\tr g}}\tr b^{-1\;ik}_{\;\;\;\;\;E}\hat L_{ik\;lm}\tr b^{-1\;lm}_{\;\;\;\;\;C}.
\nom}
To complete the algebra construction, it is also necessary to find the action of the constraint ${\cal\tilde H}^0_\xi$ on
itself. After one more cumbersome calculation we obtain:
\disn{f4}{
\pua{{\cal\tilde H}^0_\xi}{{\cal\tilde H}^0_\dz}=\int d^3x\biggl(-\ls{\tilde\Psi}^l-\tr g^{lm}{\tilde\Phi}_m\rs\ls\xi \tr \na_l\dz-\dz\tr \na_l\xi\rs+\ns
+\tilde\Psi^k \ls\pi^B\tilde{B}_{BA}\xi \tr \na_k \ls\pi^C{\tilde{B}^A_C}\dz\rs-
\pi^B\tilde{B}_{BA}\dz\tr \na_k \ls\pi^C{\tilde{B}^A_C}\xi\rs\rs\biggr).
\nom}
Thus the system of constraints turns out to be closed: the Poisson brackets of the constraints with each
other turn out to be proportional to the constraints themselves. Eqs. (\ref{g5}), (\ref{q4.1}), (\ref{g5.2}), (\ref{f2}), (\ref{l107}) and (\ref{f4}) represent
the obtained algebra of constraints in the Regge-Teitelboim formulation of gravitation
using the "external" time, i.e., under partial gauge fixing (\ref{2}) in the action. It is useful to note that the
Poisson brackets of the three constraints $\tilde\Psi^i$ with all constraints turn out to be proportional to themselves,
i.e., they form a subalgebra which is an ideal.

\section{Comparison with the case of gauge fixing\\ in the Hamiltonian formalism}
Let us check whether the canonical formalism obtained in the previous sections for the Regge-Teitelboim
formulation of gravitation using the "external" time is reproduced if one introduces partial gauge
fixing (\ref{2}) in the form of an additional condition in the canonical description of the theory with arbitrary
time. This question requires a separate study because, unlike the examples considered in Section 2,
a certain part of the constraints considered in the present approach (the Einsteinian constraints (\ref{l60})) do not follow from the action but are imposed additionally.

Let us write down the Hamiltonian of the theory with arbitrary time and the corresponding algebra
of constraints, obtained in \cite{statja24} (some misprints have been corrected in the relations presented). The
Hamiltonian has the form of a linear combination
\disn{g14}{
H^{\text{gen}}=\int\! d^3x\ls\la^i\Phi_i+N_i\Psi^i+N_4\Psi^4+N_0{\cal H}^0\rs
\nom}
of eight constraints:
\disn{t27}{
\Phi_i=\pi_a\tr{e}^a_i,
\qquad
\Psi^i=-\sqrt{-\tr{g}}\;\tr{\na}_k\!\ls\frac{1}{\sqrt{-\tr{g}}}\;\pi^a\al_a^{ik}\rs+\pi^a\tr{e}_a^i,
\no
\Psi^4=\pi_a w^a,
\qquad
{\cal H}^0=\frac{1}{4\sqrt{-\tr{g}}}\;\pi^a\al_a^{ik}\hat L_{ik,lm}\al^{lm}_b\pi^b-
\sqrt{-\tr{g}}\;\tr b^a_{ik}\tr b^b_{lm}\eta_{ab}L^{ik,lm},
\nom}
where
\disn{t20}{
w_a\tr{e}^a_l=0,\qquad w_a\tr{b}^a_{ik}=0,\qquad |w_a w^a|=1,
\nom}\vskip -2em
\disn{t20.1}{
\al^{ik}_a=\al^{ki}_a,\quad
\al^{ik}_a\, \tr{e}^a_l=0,\quad
\al^{ik}_a w^a=0,\quad
\al^{ik}_a\, \tr{b}^a_{lm}=\frac{1}{2}\ls\de^i_l\de^k_m+\de^i_m\de^k_l\rs.
\nom}
The algebra of constraints, written in terms of the convolutions
\disn{g3}{
\Phi_\xi\equiv\int d^3x\; \Phi_i(x)\,\xi^i(x),\qquad
{\cal H}^0_\xi\equiv\int d^3x\; {\cal H}^0(x)\,\xi(x),\no
\Psi_\xi\equiv\int d^3x\ls \Psi^i(x)\,\xi_i(x)+\Psi^4(x)\,\xi_4(x)\rs,
\nom}
has the form \cite{statja24}:
 \disn{m90}{
\pua{\Phi_\xi}{\Phi_\dz}=-\int d^3x\;\Phi_k\ls\xi^i\tr{\na}_i\dz^k-\dz^i\tr{\na}_i\xi^k\rs,
\nom}\vskip -1em
 \disn{m91}{
\pua{\Phi_\xi}{\Psi_\dz}=-\int d^3x\ls\Psi^k\ls\xi^i\tr{\na}_i\dz_k+\dz_i\tr{\na}_k\xi^i\rs+\Psi^4\,\xi^i\dd_i\dz_4\rs,
\nom}\vskip -1em
 \disn{m92}{
\pua{\Phi_\xi}{{\cal H}^0_\dz}=-\int d^3x\;{\cal H}^0\,\xi^i\dd_i\dz,
\nom}\vskip -1em
\disn{m93}{
\pua{\Psi_\xi}{\Psi_\dz}=\int d^3x\;\ls\de y^a_{\Psi_\xi} \,\overline\Psi_{ab}\;\de y^b_{\Psi_\dz}-
\de y^a_{\Psi_\dz} \,\overline\Psi_{ab}\;\de y^b_{\Psi_\xi}\rs,
\nom}\vskip -1em
 \disn{m94}{
\pua{\Psi_\xi}{{\cal H}^0_\dz}=\int d^3x\;\ls\de y^a_{\Psi_\xi} \,\overline\Psi_{ab}\;\de y^b_{{\cal H}^0_\dz}-
\de y^a_{{\cal H}^0_\dz} \,\overline\Psi_{ab}\;\de y^b_{\Psi_\xi}\rs,
\nom}\vskip -1em
 \disn{m95}{
\pua{{\cal H}^0_\xi}{{\cal H}^0_\dz}=
\!\int\! d^3x\Biggl(
\de y^a_{{\cal H}^0_\xi}\,\overline\Psi_{ab}\;\de y^b_{{\cal H}^0_\dz}\,-\,
\de y^a_{{\cal H}^0_\dz}\,\overline\Psi_{ab}\;\de y^b_{{\cal H}^0_\xi}\,-\,
\ls\Psi^k-\tr{g}^{kl}\Phi_l\rs\!\ls\xi\tr{\na}_k\dz-\dz\tr{\na}_k\xi\rs\!\Biggr).
\nom}
Here we have used the differential operator
 \disn{g10.1}{
\overline\Psi_{ab}=\ls\Psi^i\eta_{ab}-\Psi^4\frac{w_b}{w_c w^c} \ls\al^{ik}_a\tr{\na}_k+\tr{e}^i_a\rs\rs \tr{\na}_i,
\nom}
as well as the notations
 \disn{g10.2}{
\de y^a_{\Psi_\xi}(x)=\pua{\Psi_\xi}{y^a(x)}=
\al^{ika}\tr{\na}_i \xi_k+\tr{e}^{ia} \xi_i+w^a\xi_4,\qquad
\de y^a_{{\cal H}^0_\dz}(x)=\pua{{\cal H}^0_\dz}{y^a(x)}=\hat B^{ac}\pi_c\dz,
\nom}
where
 \disn{g11}{
\hat B^{ac}=\frac{1}{2\sqrt{-\tr{g}}}\;\al^a_{ik}\al^c_{lm}\hat L^{ik,lm}.
\nom}

In a similar manner to the way it was done in the examples of Section 2, we perform partial gauge
fixing (\ref{2}), introducing it as the additional condition $\chi=y^0-x^0\approx0$.
It is easy to notice that the Poisson bracket between this condition and one of the
constraints has the form
\disn{q6.1}{
\pua{\Psi^4}{\chi}=w^{(0)}.
\nom}
As has been mentioned at the beginning of Section~4, if the condition (\ref{2}) is satisfied, the 3-dimensiona submanifold $x^0=const$ turns out to be embedded in the 9-dimensional flat space $y^{0}=const$, whence it follows that
$\tr{e}^{(0)}_i=0$, $\tr{b}^{(0)}_{ik}=0$. In this case from (\ref{t20}) in the generic situation it follows
\disn{q6.2}{
w^A=0,\qquad |w^{(0)}|=1.
\nom}
Owing to that, one can conclude from (\ref{q6.1}) that $\Psi^4$ and $\chi$ form a pair of second-class constraints, and one
can try to solve them by expressing in their terms the pair of conjugated variables $y^0,\pi_0$.
With (\ref{q6.2}), turning to zero of the constraint $\Psi^4$ means that $\pi_0=0$. Since
the generalized momentum is equal to zero, a substitution of the values of variables to be eliminated, found
by solving the constraints $\chi,\Psi^4$, to the first-order Lagrangian is equivalent to their substitution directly
to the generalized Hamiltonian (\ref{g14}). Such a situation takes place, in particular, in the example from Section~2.1, whereas in the example from Section~2.2, for which after solving the constraint the momentum
turns out to be nonzero, a direct substitution to the generalized Hamiltonian would give, as is easily seen, a wrong result.

It is easy to verify that, as a result of a substitution of the values of $y^0=x^0$, $\pi_0=0$ found from the constraints $\chi,\Psi^4$, to the generalized Hamiltonian (\ref{g14}) one obtains the expression for the Hamiltonian in
the form (\ref{l25.1}). To do that, one should notice, by comparing Eqs.~(\ref{t27}) and (\ref{l25}), that the constraints $\Phi_i,\Psi^k,{\cal H}^0$ turn into the constraints $\tilde\Phi_i,\tilde\Psi^k,{\cal\tilde H}^0$, respectively. At the comparison, it is necessary to take into account that if $y^0=x^0$, then for the quantity $\al^{ik}_a$ determined from (\ref{t20.1}) one can write
\disn{q6.3}{
\al^{ik}_{(0)}=0,\qquad
\al^{ik}_A=\tr b^{-1\;ik}_{\;\;\;\;\;A},
\nom}
see~(\ref{l26}).

Further, one can verify that the algebra of constraints (\ref{m90})-(\ref{m95})
of the theory with arbitrary time,
after explicitly solving the constraints $\chi,\Psi^4$
turns
into the algebra of constraints
(\ref{g5}), (\ref{q4.1}), (\ref{g5.2}), (\ref{f2}), (\ref{l107}), (\ref{f4}) for the canonical formulation of the theory
with "external" time, obtained in Section 5. To do
that, it is necessary to use the relations which follow
from a comparison of Eqs.
(\ref{g10.2}) and (\ref{l103}):
\disn{p10.2}{
\de y^A_{\Psi_\xi}=\tr b^{-1\;ik\;A}\tr \na_i\xi_k+\tr e^{iA}\xi_i=r_{\xi}^A,\qquad
\de y^A_{{\cal H}^0_\dz}=\hat B^{AC}\pi_C\dz=\tilde B^{AC}\pi_C\dz.
\nom}

Thus, despite the emergence of some constraints {\it ad hoc} in the canonical formalism, a transition to
the "external" time by introducing the additional condition (\ref{2}) to the canonical description of Regge-
Teitelboim gravitation gives the same result as usage of the "external" time directly in the action, as has
been done in Sections 4 and 5. In \cite{statja24} it has been shown that if one considers the system dynamics
under the fulfilled constraints $\Psi^i$ and $\Psi^4$, then it reduces to GR dynamics in the ADM formalism.
Since the transition to "external" time corresponds to solving the constraint $\Psi^4$, it follows that the external time
dynamics of the Regge-Teitelboim formulation of gravity described in Sections~4 and 5, with the
fulfilled three constraints $\Psi^i$, also reduces to GR dynamics.

As a result of the above analysis of the equations of motion and the canonical structure of the
Regge-Teitelboim formulation of gravitation, one can conclude that a transition from the internal time $x^0$ to
the "external" time $y^0$ in the framework of a classical description does not change the physical content of
the theory, corresponding to GR. The opportunity of working with "external" time can be further used in
attempts to quantize the theory and at studies of the splitting theory \cite{statja25}, mentioned in the introduction,
which is a variant of embedding theory which does not require the usage of space-time coordinates.

\vskip 0.5em
\textbf{Acknowledgments}.
The authors thank A.N.~Semenova for useful discussions.
The work was partially supported by the Saint Petersburg State University grant N~11.38.660.2013 (S.A.P.).


\end{document}